\documentclass[twocolumn,showpacs,preprintnumbers,prb,fleqn]{revtex4}
\usepackage{epsfig}
\usepackage{graphicx}
\usepackage{amsfonts}

\newcommand{\ct}{\cite}
\newcommand{\bi}{\bibitem}
\newcommand{\be}{\begin{equation}}
\newcommand{\ee}{\end{equation}}
\newcommand{\ba}{\begin{eqnarray}}
\newcommand{\ea}{\end{eqnarray}}
\newcommand{\non}{\nonumber}

\newcommand{\ket}[1]{|#1\rangle}

\begin{document}

\title{Quenching Dynamics of a quantum $XY$ spin-1/2 chain in presence of 
a transverse field}
\author{Victor Mukherjee$^1$, Uma Divakaran$^{2}$, Amit Dutta$^3$, and 
Diptiman Sen$^4$}
\affiliation{$^{1,2,3}$Department of Physics, Indian Institute of Technology 
Kanpur 208016, India \\
$^4$Center for High Energy Physics, Indian Institute of Science, Bangalore 
560012, India}

\begin{abstract}
We study the quantum dynamics of a one-dimensional spin-1/2 anisotropic 
$XY$ model in a transverse field when the transverse field or the anisotropic 
interaction is quenched at a slow but uniform rate. The two quenching schemes 
are called transverse and anisotropic quenching respectively.
Our emphasis in this paper is on the anisotropic quenching scheme and we 
compare the results with those of the other scheme. In the process of 
anisotropic quenching, the system crosses all the quantum critical lines of 
the phase diagram where the relaxation time diverges. The evolution is 
non-adiabatic in the time interval when the parameters are close to their 
critical values, and is adiabatic otherwise. The density of defects produced 
due to non-adiabatic transitions is calculated by mapping the many-particle 
system to an equivalent Landau-Zener problem and is generally found to vary as
$1/\sqrt{\tau}$, where $\tau$ is the characteristic time scale of quenching, a
scenario that supports the Kibble-Zurek mechanism. Interestingly, in the case 
of anisotropic quenching, there exists an additional non-adiabatic transition,
in comparison to the transverse quenching case, with the corresponding 
probability peaking at an incommensurate value of the wave vector. In the 
special case in which the system passes through a multi-critical point, the 
defect density is found to vary as $1/\tau^{1/6}$. The von Neumann entropy of 
the final state is shown to maximize at a quenching rate around which the 
ordering of the final state changes from antiferromagnetic to ferromagnetic. 
\end{abstract}

\maketitle

\section{Introduction}

A quantum phase transition corresponds to a fundamental change in the
symmetry of the ground state of
a quantum system when the strength of the quantum fluctuations is 
appropriately tuned at zero temperature \ct{sachdev99,dutta96}. In a quantum
system, statics and dynamics are intermingled, and a quantum critical point 
is therefore associated with a diverging correlation length as well as a 
diverging relaxation time. The diverging time scale plays a non-trivial role 
when the system is driven through the quantum critical point at a uniform rate
\ct{zurek05,dziarmaga05,damski05,levitov06,polkovnikov05}. This implies that 
no matter how slow the quenching may be, the dynamics of the system fails to 
be completely adiabatic when a quantum critical point is crossed. The 
possibility of experimental studies of non-equilibrium strongly correlated 
quantum systems \ct{expts} has paved the way for rigorous theoretical 
investigations \ct{sengupta04,calabrese05,das06,cincio07,cramer07} of the 
dynamics of various model Hamiltonians when swept through their quantum 
critical points. Recently a general analysis has been carried out of the 
effects of quenching in gapless systems \ct{polkovnikov07}.

The above mentioned non-adiabaticity arising due to the diverging relaxation
time near the critical point leads to the production of topological defects.
The spatial distribution of the spins in the final state is quite complex 
in comparison to the situation in which the 
dynamics is adiabatic for the entire range of time. The rate of production of 
defects can be quantified using the prediction of the Kibble-Zurek (KZ) theory
extended to quantum spin chains \ct{kibble76,zurek85}. If the gap of a 
one-dimensional quantum Hamiltonian is changed linearly as $t/\tau$, where 
$\tau$ is the characteristic time scale of quenching, the dynamics is adiabatic
for almost the entire span of time, except for a region in the vicinity of the
quantum critical point called the ``impulse" region where non-adiabatic 
transitions dominate \ct{zurek05}. The system enters the impulse region at a 
time $\tilde t$ when the rate of change of the gap is of the order of the 
relaxation time of the system. One can show that $\tilde t \sim \sqrt{\tau}$ 
which eventually results in the density of defects decreasing as 
$1/\sqrt{\tau}$. In other words, there is, on average, a single defect in 
a region of length $\tilde \xi$ which also scales as $\sqrt{\tau}$. A smaller 
quenching rate (larger value of $\tau$) yields a larger $\tilde \xi$ and the 
defect production is less. A detailed analysis of the above mechanism for 
exactly solvable quantum spin chains is presented in Refs. 
\onlinecite{zurek05,dziarmaga05,damski05,levitov06,polkovnikov05} and 
\onlinecite{damski06}. Recently, the above studies have been extended to 
explore the dynamics of disordered transverse Ising chains 
\ct{dziarmaga06,caneva07}. The quenching behavior of Bose-Hubbard models has 
also been explored in recent years \ct{cucchietti07,schutzhold06}.

There is a recent upsurge in the study of quantum dynamics from the point of 
view of an optimization problem where the strength of the quantum fluctuations
is quenched from a very high value to zero in order to arrive at the true 
ground state of a frustrated classical system. This approach of adiabatic
quantum computation is popularly known as ``quantum annealing" 
\ct{kadowaki98,arnab05,sei05}. The measure of non-adiabaticity in this approach
is given by the residual energy $E_{\rm res} = E_{\rm fin} - E_{\rm cl}$,
where $E_{\rm fin}$ is the energy of the final state, and $E_{\rm cl}$ is the 
energy of the true classical ground state. The two measures of the degree of 
non-adiabaticity, the density of defects in the previous approach and the
residual energy in the quantum annealing approach, are proportional to 
each other for a disorder free system. In the present literature on quantum 
dynamics, the terms ``quenching" and ``annealing" are often used synonymously.

In this work, we study the dynamics of an anisotropic transverse $XY$ spin-1/2
chain \ct{lieb61,barouch71,bunder99} which is driven across various quantum 
critical lines at a steady and finite rate. In the transverse quenching 
scheme, the transverse field is varied from $-\infty$ to $\infty$ 
\ct{levitov06,dziarmaga05,polkovnikov05}, whereas in the ``anisotropic 
quenching", which 
constitutes the main theme of this paper, the interaction term is quenched
from $-\infty$ to $\infty$ keeping the transverse field unchanged. We study 
the ``anisotropic" quenching scheme in detail, and compare the results with 
those of the transverse quenching scheme. In both cases, the dynamics is 
exactly solved via a mapping to an equivalent Landau-Zener problem \ct{landau}
through a Jordan-Wigner transformation from spins to fermions \ct{lieb61}.

The paper is organized in the following way. In Section II, we discuss the
Hamiltonian and the corresponding phase diagram. For the sake of completeness
and to fix our notations and terminology at the outset, we include a brief
discussion of the Jordan-Wigner transformation and the diagonalization of the
spin Hamiltonian in the fermionic representation. We report the results of 
the transverse as well as the anisotropic quenching scheme in Section III. In 
Section IV, we discuss the behavior of the von Neumann entropy and the 
magnetization of the final state as a function of the quenching time.

\section{The Model and the phase diagram}

The Hamiltonian of the one-dimensional anisotropic spin-1/2 $XY$ chain in
a transverse field is given by \ct{lieb61,barouch71,bunder99} 
\be H ~=~ - \frac{1}{2} ~\sum_n ~(J_x \sigma^x_n \sigma^x_{n+1} + 
J_y \sigma^y_n \sigma^y_{n+1} + h \sigma^z_n), \label{h1} \ee
where the $\sigma$'s are Pauli spin matrices satisfying the usual 
commutation relations. The strength of the transverse field is denoted by 
$h$, and $J_x -J_y$ is the measure of the anisotropy of interactions in the 
$x$ and $y$ directions. In this work, $J_x$, $J_y$ and $h$ are chosen to be 
non-random. To explore the excitation spectrum of the Hamiltonian, we 
choose time-independent values of the parameters in this introductory section.
In the limit $J_y =0$, Hamiltonian in Eq. (\ref{h1}) reduces to the transverse
Ising Model \ct{pfeuty70}, while for $J_x = J_y$ it describes an isotropic
$XY$ model \ct{lieb61}.

The Hamiltonian in Eq. (\ref{h1}) can be exactly diagonalized using the 
Jordan-Wigner transformation which maps a system of spin-1/2's to a system 
of spinless fermions \ct{lieb61,kogut79,bunder99}. The Jordan-Wigner 
transformation of spins to fermions is given by
\ba c_n &=& \left( \prod_{j=-\infty}^{n-1} \sigma_j^z \right) ~(-1)^n ~
\sigma_n^- , \non \\ 
c_n^{\dagger} &=& \left( \prod_{j=-\infty}^{n-1} \sigma_j^z \right) ~(-1)^n ~
\sigma_n^+ , \ea
where $\sigma_n^{\pm} = (\sigma_n^x \pm i \sigma_n^y)/2$ are the spin raising 
(lowering) operators. The operator $\sigma_n^z$ is expressed in terms of 
fermion operators as $\sigma_n^z = 2 c_n^{\dagger}c_n - 1$; thus the presence 
of a fermion at site $n$ corresponds to a spin-up state. In the fermionic 
language, the above Hamiltonian can be rewritten in Fourier space with a 
periodic boundary condition as
\ba H &=& - ~\sum_{k>0} ~\{ ~[(J_x + J_y) \cos k +h] ~(c_k^{\dagger} c_k +
c_{-k}^{\dagger} c_{-k}) \non \\
& & ~~~~~~~~~~~~~~+ i (J_x -J_y) \sin k ~(c_k^{\dagger} c_{-k}^{\dagger}
- c_{-k} c_k \} . \label{h2} \ea
This Hamiltonian is quadratic in the $c$ operators and can therefore be 
diagonalized using the standard Bogoliubov transformation; we then arrive 
at an expression for the gap in the excitation spectrum given by 
\ct{lieb61,bunder99}
\be \epsilon_k = [h^2 +J_x^2 + J_y^2 + 2 h (J_x + J_y) \cos k + 2 J_x J_y
\cos 2k ]^{1/2}. \label{ek} \ee

The gap given in Eq. (\ref{ek}) vanishes at $h = \mp (J_x + J_y)$ for wave 
vectors $k =0$ and $\pi$ respectively; it turns out that this signals a 
quantum phase transition from a quantum paramagnetic phase to a 
ferromagnetically ordered phase in which a discrete $Z_2$ symmetry of the 
Hamiltonian in Eq. (\ref{h1}) ($\sigma_n^x \to - \sigma_n^x$, $\sigma_n^y \to 
- \sigma_n^y$ and $\sigma_n^z \to \sigma_n^z$ for all $n$) is spontaneously 
broken. This transition belongs to the universality class of the transverse 
Ising model \ct{bunder99,pfeuty70} and is therefore referred to as the ``Ising"
transition. The spectrum is also gapless in the limit when the anisotropy
vanishes, $J_x \to J_y$, provided that $|h/2J_x| \le 1$. The line $J_x= J_y$ 
marks the phase boundary between two ferromagnetically ordered phases denoted 
by $\rm FM_x$ and $\rm FM_y$, and the corresponding phase transition is called
the ``anisotropic transition". In the $\rm FM_x$ phase, $J_x >J_y$ and hence 
the ferromagnetic ordering is in the $x$ direction, while it is the other way 
around in the $\rm FM_y$ phase. One can also check that the long-range order
in the $\rm FM_x$ or $\rm FM_y$ phase only exists for a relatively weak 
transverse field lying in the range $-J_x - J_y < h < J_x + J_y$. Each 
ferromagnetically ordered phase is further divided into a commensurate and 
an incommensurate region with the incommensurate wave vector $k_0$ given by
\be \cos k_0 ~=~ - ~\frac{h(J_x + J_y)}{4 J_x J_y}. \ee
On the anisotropic phase boundary $J_x = J_y$, the incommensurate wave vector
therefore has a value $k_0 = \cos^{-1}(-h/2J_x)$. The boundary between these 
two regions inside a ferromagnetic phase is given by the relation $h/(J_x + 
J_y) = \pm (1 - \gamma^2)$, with $\gamma \equiv (J_x -J_y)/ (J_x + J_y)$, as 
shown by the thick dashed lines in Fig. 1. The exponents associated with the 
anisotropic transition are different from the Ising case and are identical to
the exponents of a pair of decoupled Ising models \ct{barouch71,bunder99}.

We note that the points corresponding to $J_x = J_y$ and $h = \pm 2 J_y$ are
multi-critical points since more than one phase boundary passes through them.
We will see later that the quenching dynamics shows a different behavior when
the system passes through those points.

\begin{figure}[htb]
\includegraphics[height=3.1in,width=3.4in]{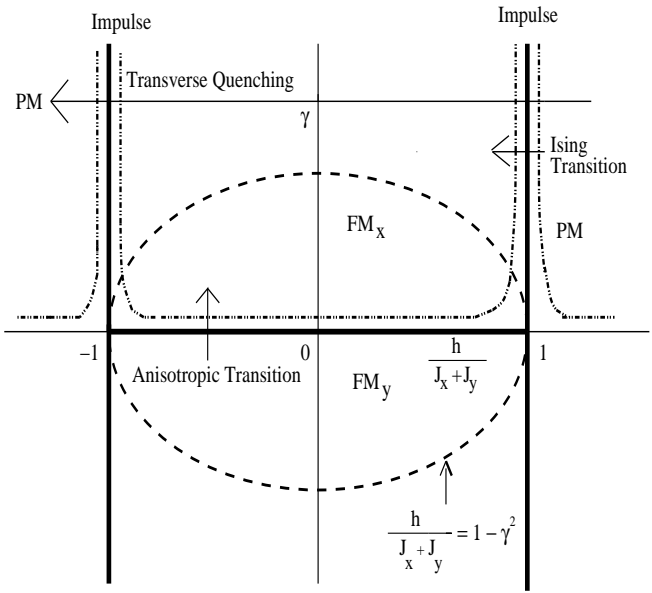}
\caption{The phase diagram of the anisotropic $XY$ model in a transverse 
field in the $~h/(J_x + J_y) ~- ~\gamma~$ plane. The vertical bold lines given 
by $~h/(J_x + J_y) = \pm 1~$ denote the Ising transitions. The system is also 
gapless on the horizontal bold line $J_x = J_y$ for $~|h| < J_x + J_y$.
${\rm FM_x ~(FM_y)}$ is a long-range ordered phase with ferromagnetic ordering
in the $x ~(y)$ direction. The thick dashed line marks the boundary between the
commensurate and incommensurate ferromagnetic phase. The thin dotted lines 
indicate the adiabatic and impulse regions when the field $h$ is quenched 
from $-\infty$ to $\infty$.} 
\end{figure}

\section {Quenching Scheme and Results}

In this section, we will discuss the results obtained for the quenching 
dynamics of the Hamiltonian in Eq. (\ref{h1}) for two different schemes. In 
the first scheme, the time-dependent transverse field is of the form $h(t) = 
t/\tau$, where $t$ is varied from $-\infty$ to $\infty$, and $\tau$ is the 
characteristic time scale of quenching, often referred to as the ``quenching 
time". This ``transverse quenching" scheme has been studied extensively for 
the $XY$ chain by Cherng and Levitov \ct{levitov06}, and for the transverse 
Ising case by Dziarmaga \ct{dziarmaga05} and by Polkovnikov \ct{polkovnikov05}. We shall briefly present the results
for the transverse quenching which will be helpful in comparing it with the 
results obtained in the other scheme, namely, the ``anisotropic quenching".

As mentioned in the Introduction, in the ``anisotropic quenching'' scheme, the
interaction term $J_x (=t/ \tau)$ is quenched from a very large negative 
initial value to a very large positive final value, with $J_y$ and $h$ held
fixed at some positive values. For our numerical studies,
we will set $2J_y > h$ for the reason explained below. In the initial 
state, all the spins are antiferromagnetically ordered in the $x$ direction. 
On the other hand, the final state should correspond to a state with a perfect
ferromagnetic order in the $x$ direction had the dynamical evolution been 
adiabatic for the entire span of time. However, due to the non-adiabatic 
transitions near the quantum critical regions, the final state in the limit 
$t \to \infty$ will not be perfectly ordered and will include a finite 
fraction of kinks or spins aligned anti-parallel to the direction of ordering.

The Hamiltonian in Eq. (\ref{h2}) decouples into a sum of independent terms, 
$H(t) = \sum_{k>0} H_k(t)$, where each $H_k(t)$ operates on a four-dimensional
Hilbert space spanned by the basis vectors $\ket 0$, $\ket k = c_k^\dagger \ket
0$, $\ket {-k} = c_{-k}^\dagger \ket 0$, and $\ket {k,-k} = c_k^\dagger 
c_{-k}^\dagger \ket 0$. The vacuum state where no $c$-particle is present is 
denoted by 
$\ket 0$ which corresponds to a spin configuration with all spins pointing in 
the $-z$ direction. The occurrence of only bilinear terms like $c_k^\dagger 
c_{-k}^\dagger$ in the Hamiltonian in (\ref{h2}) ensures that the parity (even
or odd) of the total number of fermions given by $n_k=c_k^\dagger c_k + 
c_{-k}^\dagger c_{-k}$ is conserved for each value of $k > 0$. Thus the states
$\ket 0$ and $\ket {k,-k}$ are coupled to each other by the Hamiltonian, while
the states $\ket k$ and $\ket {-k}$ remain invariant.

To study the dynamics of transverse quenching, it is sufficient to project the
Hamiltonian $H_k(t)$ to the two-dimensional subspace spanned by $\ket 0$ and 
$\ket {k,-k}$ since the ground state of Eq. (\ref{h2}) for each value of $k$
lies within this subspace. In this subspace, the Hamiltonian takes the form
\ba H_k (t) &=& -~[ h ~+~ (J_x+J_y)\cos k] ~I_2 \non \\
&+& \left[ \begin{array}{cc} h+ (J_x + J_y) \cos k & i (J_x - J_y)
\sin k \\
-i(J_x - J_y) \sin k & -h-(J_x+J_y)\cos k \end{array} \right], \non \ea
where $I_2$ denotes the $2 \times 2$ identity matrix.
A state in this subspace can be represented as a linear superposition 
$\psi_k(t) = u_k(t)|0> + v_k(t)|k,-k>$, where the amplitudes $u_k(t)$ and 
$v_k(t)$ are time-dependent. The initial condition in the transverse quenching
scheme is given by $u_k(-\infty)=1$ and $v_k(-\infty)=0$. The time 
evolution of a generic state is governed by the Schr\"odinger equation 
\be i \partial_t\psi_k(t) ~=~ H_k(t) ~\psi_k(t) . \label{schr} \ee

The projection of the Hamiltonian to the $2 \times 2$ Hilbert space has 
effectively reduced the many-body problem to the problem of a two-level 
system. The off-diagonal term of the projected Hamiltonian, $\Delta=(J_x -J_y)
\sin k$, represents the interaction between the two time-dependent levels 
$E_{1,2}=\pm [h(t)+(J_x + J_y)\cos k]$. The Schr\"odinger equation given above
is identical to the Landau-Zener problem of a two-level system, where the 
off-diagonal terms of the Hamiltonian determines the non-adiabatic transition 
probability $p_k$ for the wave vector $k$. Using the Landau-Zener transition 
formula, this transition probability of excitations at the final time is 
given by \ct{sei05}
\be p_k ~=~ e^{-2 \pi {\tilde \gamma}} , \label{pk} \ee
where $\tilde \gamma$ = $\Delta^2 /\left|\frac{d}{dt}(E_1 - E_2)\right|$.
Equivalently, $p_k$ determines the probability that the system remains in
the initial state $\ket 0$ at the final time. For the transverse quenching 
case, one can use the relation in Eq. (\ref{pk}) to obtain $p_k$ as a 
function of the anisotropy term $J_x -J_y$ and $\tau$ as \ct{levitov06}
\be p_k = e^{-\pi \tau (J_x - J_y)^2 \sin^2 k} . \ee
This leads to an expression for the density of kinks $n$ generated due to
non-adiabatic transitions,
\be n ~=~ \int^\pi_0 ~\frac{dk}{\pi} ~p_k ~\simeq ~\frac{1}{\pi \sqrt{\tau}~
|J_x - J_y|} . \label{n1} \ee
Eq. (\ref{n1}) shows that the kink density decreases as $1/\sqrt{\tau}$ for
large $\tau$ as predicted by the Kibble-Zurek theory and proved by Cherng and
Levitov in the case of transverse quenching \ct{levitov06}. 

We shall now focus on the anisotropic quenching scheme where the interaction 
term $J_x(t) = t/\tau$ is changed from $-\infty$ to $\infty$, with $J_y$ and 
$h$ held fixed at some positive values. Let us first point out the range in 
time where the system fails to follow the instantaneous ground state (namely,
the region where the relaxation time is large or divergent) as $J_x$ is 
varied. If $J_x (t_1) + J_y =-h$, i.e., for $t_1 = -\tau (h+J_y)$, the system 
undergoes an Ising transition from the initial antiferromagnetic phase to a 
paramagnetic phase. When $J_x$ is further increased so that $J_x (t_2) + J_y 
= h$, i.e., for $t_2 = (h - J_y)\tau $, there is a phase transition from the 
paramagnetic to $\rm FM_y$ phase. It is to be noted however, that for
$J_x = -J_y$, there is no further anisotropic transition since the magnitude
of $h$ is greater than $J_x + J_y = 0$, and the system stays paramagnetic. 
Eventually, at a time $t_3$ given by $J_x (t_3) = J_y$, with $h < 2J_y$, the
system evolves to the $\rm FM_x$ phase. The non-adiabaticity dominates in the 
vicinity of these three quantum critical points, i.e., for $t$ close to 
$t_1,t_2$ and $t_3$. One should also note that if we fix $h > 2J_y$ at the 
outset, there cannot be any anisotropic transition, and the system directly 
evolves from the paramagnetic phase to the $\rm FM_x$ phase through the Ising
transition only; hence there are two, rather than three, regions of 
non-adiabaticity. All the transition points are depicted in the static phase 
diagram of the model shown in Fig. 1. The adiabatic and non-adiabatic regions 
of time evolution are shown schematically in Fig. 2.
 
\begin{figure}[htb]
\includegraphics[height=2.5in,width=3.6in,angle=0]{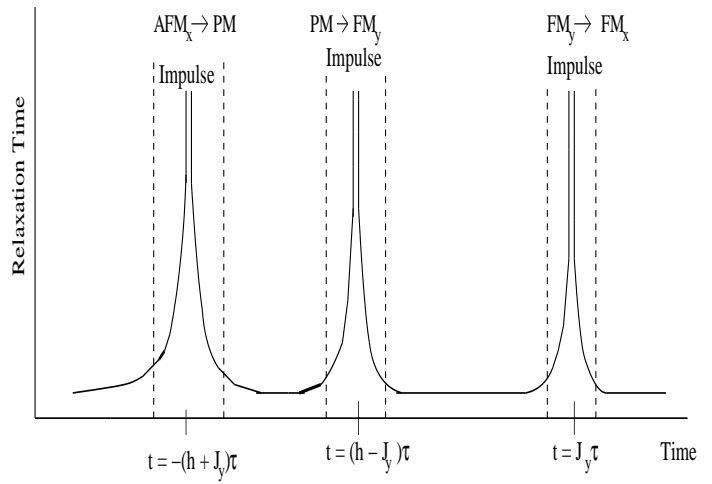}
\caption{Schematic diagram showing the divergence of the relaxation time of 
the quantum Hamiltonian at the quantum critical points. As discussed in the 
text, the dotted vertical lines denote the impulse region of dynamics where 
non-adiabatic transitions play a prominent role.}
\end{figure}

As mentioned already, in the anisotropic case,
the reduced Hamiltonian in the $\ket 0, ~\ket {k,-k}$ subspace includes
a time-dependent $J_x$ term with static positive values of $J_y$ and $h$. 
The eigenstates of the Hamiltonian in the limit $t \to \pm \infty$ are
$$\ket {e_{1k}} = \sin (k/2) \ket 0 + i\cos (k/2) \ket {k,-k}$$ and 
$$\ket {e_{2k}} = \cos (k/2) \ket 0 - i\sin (k/2) \ket {k,-k},$$
with eigenvalues $\lambda_1 = t/\tau$ and $\lambda_2 = -t/\tau$ respectively; 
the system is in the state $\ket {e_{1k}}$ initially. A general state vector 
can be expressed as a linear combination of $\ket e_{1k}$ and $\ket e_{2k}$,
\be \ket {\psi_k(t)} = C_{1k}(t) \ket {e_{1k}} + C_{2k}(t) \ket {e_{2k}} . \ee
The initial condition in the anisotropic case is $C_{1k}(-\infty) = 1$ and 
$C_{2k}(-\infty) = 0$. The amplitudes $u_k$ and $v_k$ are related to the new 
coefficients as
\ba u_k &=& C_{1k} \sin (k/2) ~+~ C_{2k} \cos (k/2) , \non \\
v_k &=& iC_{1k} \cos (k/2) ~-~ iC_{2k} \sin (k/2) . \label{ukvk} \ea
The Landau-Zener transition probability can be obtained by numerically solving
the Schr\"odinger equation in Eq. (\ref{schr}) using the basis vectors 
$\ket 0$ and $\ket {k,-k}$, with the appropriate initial conditions for 
$u_k$ and $v_k$. The non-adiabatic transition probability $p_k$ at the final 
time is simply given by $p_k = |C_{1k} (\infty)|^2$. We have shown the 
variation of $p_k$ with $k$ obtained through numerical integration in Fig. 3.

\begin{figure}[htb]
\includegraphics[height=2.7in,width=3.4in,angle=0]{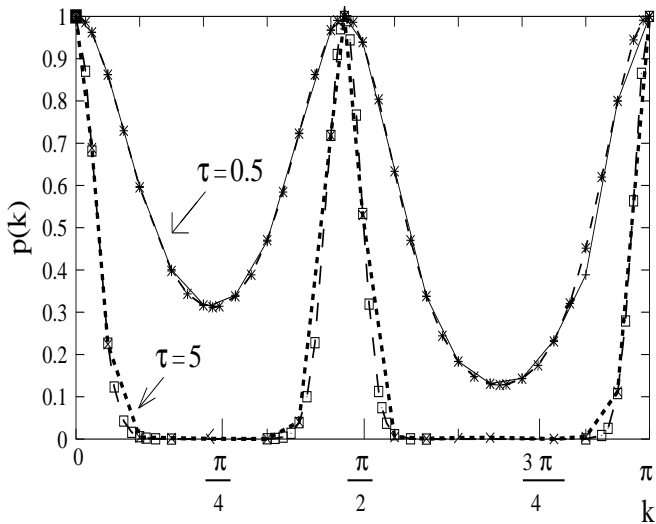}
\caption{$p_k$ vs $k$ as obtained numerically and analytically for $\tau$ 
equal to 0.5 and 5. We have fixed $J_y=1$ and $h=0.2$. In the region marked 
$\tau = 0.5$, the solid line is numerical and dashed line is analytical. In 
the region marked $\tau = 5$, the dashed line with smaller spacing is 
numerical and dashed line with larger spacing is analytical.}
\end{figure}

The situation is apparently complicated in the anisotropic quenching case
because the off-diagonal terms of the $2 \times 2$ Hamiltonian matrix in the
$\ket 0$ and $\ket {k,-k}$ basis are time-dependent. However, by an 
unitary transformation to the new set of basis vectors 
$\ket e_{1k}$ and $\ket e_{2k}$, one can get rid of the time-dependence of 
the off-diagonal terms and map the anisotropic quenching problem to an 
equivalent Landau-Zener problem. The unitary transformation is given by 
$H'_k (t) = U^\dagger H_k (t) U$, where 
\ba U ~=~ \left[ \begin{array}{cc} \cos (k/2) & \sin (k/2) \\
- i \sin (k/2) & i \cos (k/2) \end{array} \right], \non \ea
and the new Hamiltonian is
\ba H'_k (t) ~=~ -~[ h ~+~ (J_x+J_y)\cos k] ~I_2 \non \ea
\ba + \left[ \begin{array}{cc} J_x + J_y \cos 2k + h \cos k & J_y \sin 2k
+ h \sin k \\
J_y \sin 2k + h\sin k & -J_x - J_y\cos 2k - h \cos k \end{array} \right]. 
\non \ea
Therefore, the anisotropic quenching is now placed on the same footing as the 
transverse quenching case, with off-diagonal terms in the Hamiltonian matrix 
which are time-independent. The time evolution of the amplitudes $C_{1k} (t)$
and $C_{2k} (t)$ is dictated by the Schr\"odinger equation,
\ba i\frac{dC_{1k}}{dt} &=& (J_x(t) + J_y\cos 2k + h\cos k) ~C_{1k}(t) \non \\
& & + ~(J_y \sin 2k + h\sin k) ~C_{2k}(t) \non \\
i\frac{dC_{2k}}{dt} &=& (J_y \sin 2k + h\sin k) ~C_{1k}(t) \non \\
& & - ~(J_x(t) + J_y\cos 2k + h\cos k) ~C_{2k}(t), \ea
where we have dropped the effect of the term involving the identity matrix 
in the Hamiltonian since this affects $C_{1k}$ and $C_{2k}$ by the same 
time-dependent phase factor, and will therefore have no effect on the 
density matrix. The non-adiabatic transition probability now depends on 
$J_y$ and $h$, and is given by
\be p_k ~=~ |C_{1k}(\infty)|^2 ~=~ e^{-\pi \tau (J_y\sin 2k + h\sin k)^2}. \ee
In Fig. 3, we plot $p_k$ as a function of the wave vector $k$ along with
the numerical results. We see that
in contrast to the transverse case, there is an additional region of 
non-adiabaticity peaked at the incommensurate wave vector $k_0 = \cos^{-1} 
(- h/2J_y)$. This non-adiabaticity arises due to the existence of the 
anisotropic transition of the underlying static $XY$ model at $J_x(t) = J_y$.
The situation with zero transverse field is a special case where the regions 
of non-adiabaticity peak at $k = 0, \pi/2$ and $\pi$, respectively.

As in the transverse quenching case, we now measure the density of kinks 
given by
\ba n ~=~ \int^{\pi}_0 ~\frac{dk}{\pi} ~p_k ~=~ \int_0^\pi ~
\frac{dk}{\pi} ~e^{-\pi \tau (J_y\sin 2k + h \sin k)^2}. \label{npk} \ea
In Fig. 4, the plot of the kink density as a function of the quenching time 
$\tau$ is shown, which clearly shows that $n \propto 1/\sqrt{\tau}$ for large 
values of $\tau$. This finding supports the prediction of the Kibble-Zurek 
mechanism even in the anisotropic quenching case. One can also find an 
approximate analytical form of $n$ in the following way: for large $\tau$, 
only the modes very close to the critical modes contribute. For $k \to 0$ 
and $\pi$, $p_k$ can be approximated as $\exp [- \pi \tau (2J_y + h)^2 k^2]$ 
and $\exp [- \pi \tau (2J_y - h)^2 (\pi - k)^2]$ respectively. With this $p_k$,
the density of kinks produced by the modes near $k=0$ and $\pi$ is given by
\ba n_1 \simeq \frac{1}{2 \pi \sqrt{\tau}} \left[ \frac{1}{2J_y + h} + 
\frac{1}{2J_y - h} \right] = \frac{2J_y}{\pi \sqrt{\tau} (4J_y^2 - h^2)}. 
\non \ea
By expanding around $k_0$, we find that the contribution to $n$ from modes 
with $k \sim k_0$ is equal to the contribution from $k=0$ and $k=\pi$ taken 
together, 
\ba n_2 &\simeq& \int_0^\pi ~\frac{dk}{\pi} ~e^{-\pi \tau (2J_y \cos 2k_0 + 
h \cos k_0)^2 (k-k_0)^2} \non \\ 
&\simeq& \frac{1}{\pi \sqrt{\tau}} ~\frac{1}{|2J_y \cos 2k_0 +h \cos k_0|}. 
\non \ea
Therefore the total kink density is given by
\be n ~=~ n_1 + n_2 ~\simeq~ \frac{4J_y}{\pi \sqrt{\tau} ~(4J_y^2 - h^2)}. 
\label{total} \ee
Fig. 4 shows that this approximate form embraces the exact result perfectly 
in the limit of large $\tau$.

\begin{figure}[htb]
\includegraphics[height=2.5in,width=2.8in,angle=0]{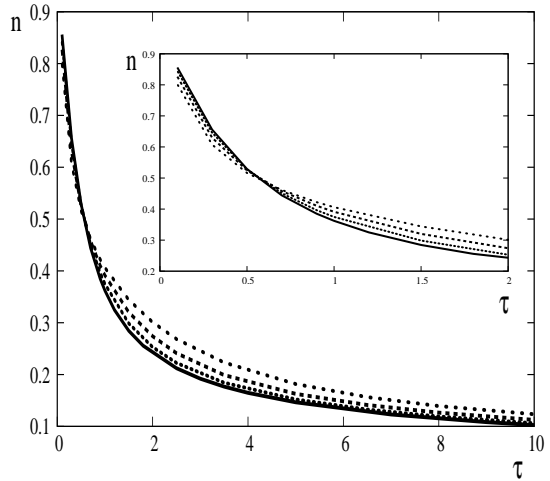}
\caption{Variation of kink density $n$ with $\tau$ as obtained by numerical
integration of Eq. (\ref{npk}) for $h = 0.2, 0.4, 0.6, 0.8$ (from bottom to 
top in the large $\tau$ region), with $J_y= 1$. For large $\tau$, $n$ 
increases with increasing $h$, whereas for small $\tau$, it decreases with 
increasing $h$ as shown in the inset.} 
\end{figure}

Using Eq. (\ref{total}), one can find the variation of the kink density with 
$h$, for large values of $\tau$, as
\ba \frac{\partial n}{\partial h} ~\simeq~ \frac{8hJ_y}{\pi \sqrt{\tau} ~
(4J_y^2 - h^2)^2}. \non \ea
The slope is positive for all values $h$ indicating an increment 
in kink density with increasing $h$ as shown in Fig. 4. On the other hand, 
for small $\tau$, the density $n$ is found to decreases with increasing $h$. 
In the limit $\pi \tau (J_y \sin 2k + h \sin k)^2 \ll 1$, one can expand the 
exponential in $p_k$ in Eq. (\ref{npk}), retaining only terms up to first 
order in $\tau$. One can then show that $\partial n/\partial h$ is negative.
The crossover from the small $\tau$ to the large $\tau$ behavior occurs around
a typical quenching time $\tau = \ln 2/[\pi (J_y \sin 2k + h \sin k)^2]$.
The significance of the crossover time is explained below.

To derive a characteristic time scale of the anisotropic quenching, we first 
note that the minima of the probability $p_k$ occurs at a wave vector value 
$\tilde k$, where 
\be \cos \tilde k ~=~ \frac{-h \pm \sqrt{h^2 + 32 J_y^2}}{8J_y}. \label{tk} \ee
The presence of a non-zero transverse field $h$ leads to an asymmetry in $p_k$
on the either side of the maxima at $k_0$ as is evident at smaller values of
$\tau$ (Fig. 3). We now define a time scale $\tau_0$ so that at $\tau = 
\tau_0$, $p_k = 1/2$ at the minima\ct{levitov06}$ k = \tilde k$. This implies
\be \tau_0 = \frac{\ln 2}{\pi(J_y \sin 2 \tilde k + h\sin \tilde k)^2}. \ee
The two conjugate values of $\tilde k$ given in Eq. (\ref{tk}) lead to a 
pair of $\tau_0$'s for the anisotropic quenching, in contrast to the transverse
quenching case. For $h=0$, these two values coalesce into one. The crossover 
of density shown in Fig. 4, takes place roughly around the smaller value of 
$\tau_0$ denoted by $\tau_{02}$. For $\tau \gg \tau_{01}, \tau_{02}$, the 
dynamics is nearly adiabatic except for the modes very close to the critical
modes (Fig. 5).

\begin{figure}[htb]
\includegraphics[height=2.7in,width=3.4in]{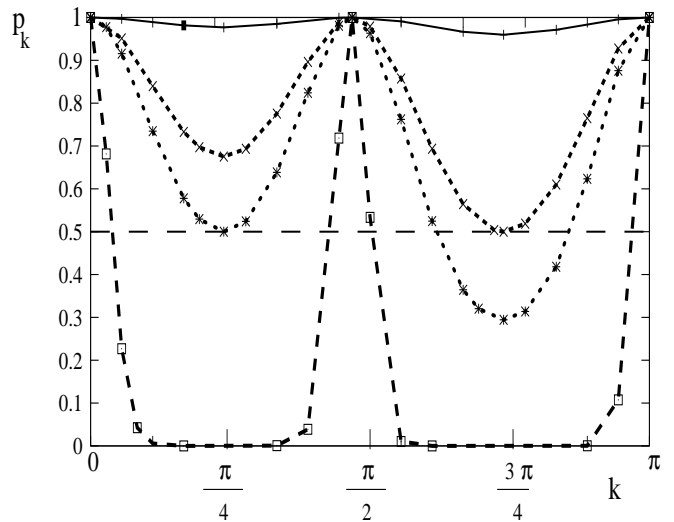}
\caption{Variation of $p_k$ vs $k$ for $\tau = 5,\tau_{01},\tau_{02}$ and 0.01
from bottom to top, where $p_k = 0.5$ at the two minima for $\tau = \tau_{01}$
and $\tau_{02}$. We have fixed $J_y = 1$ and $h = 0.2$. For $\tau \gg 
\tau_{01}, \tau_{02}$, the evolution is almost adiabatic except at $k = 0, ~
\pi$ and $k_0$, where the system remains frozen in its initial state. For 
very small $\tau$, $p_k$ is close to 1 for all $k$.}
\end{figure}

Eq. (\ref{total}) is valid for $2J_y > h$. On the other hand, if we take $h > 
2J_y$, the system evolves directly from the paramagnetic to the $\rm FM_x$
phase; there is no anisotropic transition. In this case, Eq. (\ref{npk}) gives
the kink density to be
\ba n \simeq \frac{1}{2 \pi \sqrt{\tau}} ~\left[ \frac{1}{h + 2J_y} + \frac{1}{
h - 2J_y} \right] = \frac{h}{\pi \sqrt{\tau} ~(h^2 - 4J_y^2)}. \non \ea
(Note that $\partial n/\partial h$ is negative in this case).
The transverse Ising model result given in Ref. \onlinecite{dziarmaga05} can 
be obtained by setting $J_y = 0$ in the above expression.

Finally, we consider what happens if $1-2J_y/h$ is close to zero. If $2J_y=h$,
the system passes through a multi-critical point when $J_x(t)$ equals $J_y$; 
we therefore expect an unusual behavior of the kink density $n$ in this case. 
If $2J_y = h$ and $\tau$ is large, we find that Eq. (\ref{npk}) is dominated 
by the region near $k= \pi$; since $p_k$ can be approximated there by 
$\exp [-\pi \tau h^2 (\pi - k)^6 /4]$, the contribution of this region to $n$ 
goes as $1/(h^{1/3} 
\tau^{1/6})$. [The contribution from the region near $k=0$ to $n$ goes as $1/(h
\sqrt{\tau})$ which is much smaller than $1/(h^{1/3} \tau^{1/6})$ if $h^2 \tau
\gg 1$.] Now, if $1 - 2J_y/h$ is non-zero but small, $p_k$ can be approximated
near $k=\pi$ by $\exp [-\pi \tau \{ (h- 2J_y) (\pi - k) + h (\pi - k)^3 /
2 \}^2 ]$. We then see that the term $\tau (h- 2J_y)^2 (\pi - k)^2$ will become
more important than the term $\tau h^2 (\pi - k)^6$ and the kink density will
show a crossover from a $1/\tau^{1/6}$ behavior to a $1/\sqrt{\tau}$ behavior
when $\tau$ increases beyond a value which is of order $h/|h-2J_y|^3$.

\section {Entropy and magnetization of the final state}

The quenching dynamics of the spin model is dictated by the Schr\"odinger 
equation and is obviously unitary. Therefore, the final state must be a pure 
state described by a density matrix of the product form $\rho = \bigotimes 
\rho_k$, where $\rho_k$ is given by
\ba \left[ \begin{array}{cc} p_k & q_k \\ 
q_k^* & 1 - p_k \end{array} \right] , \ea
and $q_k = C_{1k} (\infty) C_{2k}^* (\infty)$. The diagonal elements of the 
reduced $2 \times 2$ density matrix are smooth functions of $k$ and are 
independent of the total time of quenching (which we now set equal to $T$ for 
convenience, with $T \gg 1$), whereas the off-diagonal terms are rapidly 
oscillating functions of both $k$ and $T$. 

The final state, even though a pure state, has a fairly complicated local
structure; its local properties in the real space is identical to that of a 
mixed state with a finite entropy \ct{levitov06}. In other words, the final 
state can be viewed as a superposition of different configurations of 
magnetically ordered domains. The off-diagonal terms of the density matrix 
$\rho_k$ can be made to vanish at the final time $T$ upon coarse-graining in 
the wave vector $k$. The final state therefore may be viewed locally as a 
mixed state described by a decohered reduced density matrix $\rho_D$ given by
\ba \left[ \begin{array}{cc} p_k & 0 \\
0 & 1 - p_k \end{array} \right] . \ea
To quantify the amount of information lost in the decoherence process, we 
consider the von Neumann entropy density of the system, $s = -{\rm tr} 
(\rho_D \ln \rho_D)$, namely,
\ba s ~=~ -\int^\pi_0 \frac{dk}{\pi} ~[~p_k \ln (p_k) ~+~ (1 - p_k) \ln 
(1 - p_k) ~]. \ea
In the case of the transverse quenching \ct{levitov06}, the maximum of the
entropy density occurs near the value of $\tau = 2\tau_0$ where the minimum
value of $p_k = 1/2$. 
In the anisotropic case, as mentioned already, $\tau_0$ is non-unique (Fig. 5).
The entropy density $s$ when plotted against $\tau/\tau_{02}$ shows a maxima 
at a quenching rate $\tau \sim 2\tau_{02}$ (see Fig. 6), while no special 
behavior is noted near $\tau_{01}$; this establishes $\tau_{02}$ as the 
characteristic time scale of the anisotropic quenching. For sufficiently small
$\tau$, non-adiabaticity dominates and the system stays in its initial state 
with high probability, namely, the system more or less retains its initial 
antiferromagnetic ordering even at the final time. The system looks like a 
pure state even at small length scales. In the large $\tau$ limit, on the 
other hand, the system evolves essentially in an adiabatic way to the final 
ground ground state, and the non-adiabatic transition probability is small. 
The final state is ferromagnetically ordered. Therefore, the entropy density 
vanishes asymptotically for both large and small values of $\tau$, and the 
maximum occurs at an intermediate value of $\tau \sim 2\tau_{02}$ where the
defect in the local ordering of the final state is maximum.

The above arguments can be justified further by exploring the magnetization of
the system in the final state. Let us recall that the initial state is 
antiferromagnetically oriented in the $x$ direction. The average energy of the
mode $k$ at the final time is related to $p_k$ as $(2p_k - 1) J_x$; hence, 
integrating over all the modes gives the total energy per site to be 
$(2n -1)J_x$, where $n$ is the density of kinks defined above. On the other 
hand, at the final time, $J_x \gg J_y$ and $h$; hence the effective 
Hamiltonian is $H = - \sum_n J_x \sigma_n^x \sigma_{n+1}^x$, and the total 
energy density is $- J_x m_x^2$. Putting all this together and using Eq. 
(\ref{total}), we arrive at an expression for the magnetization,
\be m_x (t \to \infty) \simeq \left( 1 - \frac{8J_y}{\pi \sqrt{\tau} ~
(4J_y^2 - h^2)} \right)^{1/2}, \label{mx} \ee
and $m_x =0$ whenever the right hand side of Eq. (\ref{mx}) is imaginary. In
Fig. 6, we plot the final magnetization as a function of $\tau/\tau_{02}$
which shows that the magnetization starts to become non-zero when the quenching
time is of the order of $2 \tau_{02}$. For slower quenching (larger values of
$\tau$), the dynamics tends to be more and more adiabatic, and the final 
magnetization monotonically increases towards the saturation value of unity.

Let us now shift our attention to the behavior of the staggered magnetization 
$m_{sx}$. In the limit of small $\tau$, the system fails to follow the 
instantaneous Hamiltonian; hence the final state retains an antiferromagnetic 
order with vanishing total magnetization. In other words, the total 
magnetization of the even sites cancels the total magnetization of the odd 
sites. Using similar arguments as given above, the staggered magnetization can
be shown to behave as $(2n -1)^{1/2}$; hence it vanishes at $n=1/2$ and stays 
at zero for smaller values of $n$ (i.e., higher values of $\tau$). On the other
hand, for smaller values of $\tau$, $n>1/2$ and there is a non-zero value of 
$m_{sx}$. Fig. 6 shows that the staggered magnetization of the final state 
vanishes at a quenching rate close to $2 \tau_0$ where the ferromagnetic order
starts to set in.

We can use similar arguments as given in the case of entropy density to 
understand the variation of the magnetization with $\tau$ as presented above.
We have argued already that even though the final state is a pure state, it 
can also be viewed locally, or with a coarse-grained wave vector scale, as a
decohered (mixed) state. If the quenching time scale $\tau \ll \tau_{02}$, 
there are very few local defects and the system retains the 
initial antiferromagnetic order. On the other hand, for $\tau \gg \tau_{02}$,
the final state is locally ferromagnetically ordered. Hence, there exists an 
intermediate region of $\tau \sim \tau_{02}$ where the antiferromagnetic 
order decreases rapidly. Eventually, for $\tau \gtrsim 2\tau_{02}$, the final
state starts to acquire ferromagnetic ordering.

\begin{figure}[htb]
\includegraphics[height=3.3in,width=3.4in,angle=0]{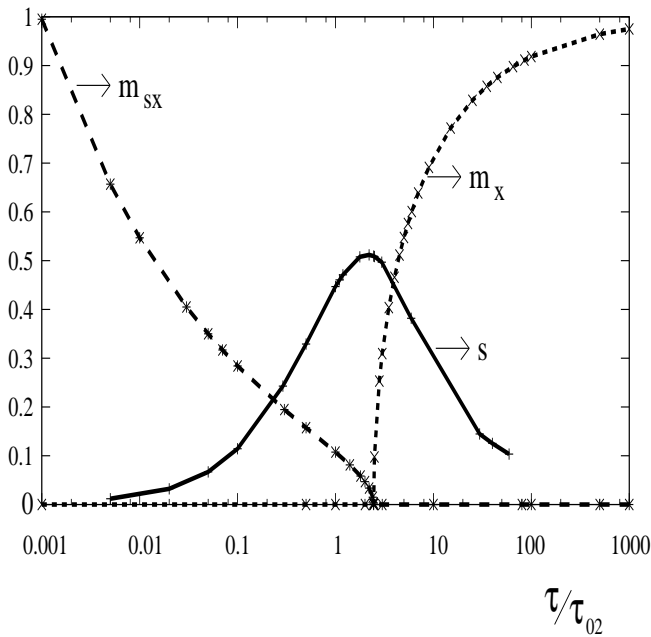}
\caption{Variation of von Neumann entropy density $s$, staggered magnetization
$m_{sx}$ and magnetization $m_{x}$ as a function of $\tau/\tau_{02}$, for 
$J_y = 1$ and $h = 0.2$.}
\end{figure}
 
\section{Conclusions}

We have studied the adiabatic quantum dynamics of an anisotropic $XY$ spin-1/2
chain
in a transverse field when the parameters of the Hamiltonian are quenched at 
a steady and finite rate. Our emphasis lies on a new scheme, namely, the 
anisotropic quenching, where the interaction strength in the $x$ direction is
quenched from $-\infty$ to $\infty$ at a uniform rate dictated by a 
time scale $\tau$. At first sight, the Landau-Zener theory does not seem to be
applicable in this case due to the presence of time-dependent off-diagonal 
terms in the reduced $2 \times 2$ Hamiltonian. However, through a unitary 
transformation to an appropriate basis, the time dependence can be entirely 
shifted to the diagonal terms, and Landau-Zener theory turns out to be 
applicable once again. 

In the process of the anisotropic quenching, the system is swept across all 
the three quantum critical lines in the phase diagram if $h < 2J_y$, and 
across two quantum critical lines if $h > 2J_y$. In both these cases, the 
density of defects is found to vary as $1/\sqrt{\tau}$. 
In the special case of $h = 2J_y$, the system is swept across a multi-critical
point, and the density of defects is found to vary as $1/\tau^{1/6}$. 

Finally, we find a characteristic quenching time $\tau_{02}$ around which the 
von Neumann entropy of the final state maximizes and ferromagnetic ordering 
starts to set in. The residual energy of the final state is proportional to 
the defect density and hence scales as $1/\sqrt{\tau}$.

\begin{center}
{\bf Acknowledgments}
\end{center}
A.D. acknowledges J. K. Bhattacharjee, B. K. Chakrabarti, A. Das and K. 
Sengupta for interesting discussions. D.S. thanks DST, India for financial 
support under the project SP/S2/CMP-27/2006.

\noindent E-mail: $^1$victor@iitk.ac.in \\ 
$^2$udiva@iitk.ac.in \\
$^3$dutta@iitk.ac.in,\\
$^4$diptiman@cts.iisc.ernet.in

\end{document}